\documentclass[a4paper]{article}

\usepackage{amsmath}
\usepackage{graphicx}   
\usepackage[american]{babel}
\usepackage[utf8]{inputenc}
\usepackage[T1]{fontenc}
\usepackage[babel=true]{csquotes}
\usepackage[backend=biber,style=ieee, sortcites]{biblatex}
\usepackage{afterpage}
\usepackage{color}		
\usepackage{epsfig}
\usepackage{authblk}
\usepackage{pdfpages}
\usepackage{graphicx,epstopdf}
\usepackage[font={footnotesize}]{caption}
\usepackage{lineno}

\graphicspath{ {Images/}}

\def\hyp{{\hbox{-}}}
\begin{document}

\title{Evolutionary advantage of directional symmetry breaking in self-replicating polymers}

\author{Hemachander Subramanian$^{1\ast}$, Robert A. Gatenby$^{1,2}$ \\
\normalsize{$^{1}$ Integrated Mathematical Oncology Department,} \\
\normalsize{$^{2}$Cancer Biology and Evolution Program, }\\
\normalsize{H. Lee Moffitt Cancer Center and Research Institute, Tampa, Florida.} \\
\normalsize{$^\ast$To whom correspondence should be addressed; E-mail:  hemachander.subramanian@moffitt.org.} }
\date{\today}
\maketitle

\begin{abstract}
Due to the asymmetric nature of the nucleotides, the extant informational biomolecule, DNA, is constrained to replicate unidirectionally on a template. As a product of molecular evolution that sought to maximize replicative potential, DNA's unidirectional replication poses a mystery since symmetric bidirectional self-replicators obviously would replicate faster than unidirectional self-replicators and hence would have been evolutionarily more successful. Here we carefully examine the physico-chemical requirements for evolutionarily successful primordial self-replicators and theoretically show that at low monomer concentrations that possibly prevailed in the primordial oceans, asymmetric unidirectional self-replicators would have an evolutionary advantage over bidirectional self-replicators. The competing requirements of low and high kinetic barriers for formation and long lifetime of inter-strand bonds respectively are simultaneously satisfied through asymmetric kinetic influence of inter-strand bonds, resulting in evolutionarily successful unidirectional self-replicators.  
\end{abstract}


\section*{Introduction}
The mechanism of replication of DNA, the universal genetic material of living systems, is far from simple. The two anti-parallel strands of a duplex DNA function as templates for the construction of daughter strands, resulting in two duplex DNA strands. But since the construction of daughter strand happens unidirectionally, from $3'$-end of the template strand towards the $5'$-end, and since the template strands are anti-parallel, one of the daughter strands, the leading strand, is constructed continuously, whereas the other lagging strand is constructed in fragments which are subsequently rejoined. Being a product of molcular evolution\cite{primitiveearth1,primitivegeneticpolymers,rnaevolved,rnaevolved2}, it would be natural to expect evolution to choose monomers supporting bidirectional replication and parallel duplex strand orientation for faster replication and to avoid the inherently complicated lagging strand replication mechanism. This leads us to question the evolutionary reasons for the choice of a) unidirectional construction of daughter strand and b) anti-parallel DNA strand orientation. 

In this article, we examine the first of the above two questions and provide a theoretical justification for the evolutionary choice of unidirectional replica strand construction over bidirectional construction. We begin by considering primordial, non-enzymatically self-replicating polymers, that evolutionarily preceded RNA and DNA. We set the stage for evolutionary competition by imagining multiple species of autocatalytic polymers, constructed out of chemically-distinct monomers, competing for common precursors, energetic sources for activation, catalytic surfaces and niches, in the primordial oceans. Our central premise is that the simplest of the evolutionary strategies, higher rates of replication\cite{lifevsprelife}, determined the outcome of this evolutionary competition. We identify some fundamental, common-sense functional requirements that these primordial autocatalytic polymers must satisfy in order to replicate faster than other competing species and hence be evolutionarily successful.

Evidently, the evolutionary search for the perfect non-enzymatically self-replicating molecular species in a given environment is constrained by the diversity of molecules available to be used as monomers in that environment, in the primordial oceans. But, this constraint is intractable, in the absence of well-established knowledge of the chemistry of primordial oceans. We circumvent this biochemical constraint by ignoring its existence, and thus \textit{theoretically assume that evolution was allowed to experiment with an infinite variety of molecular species in its search for the perfectly-adapted monomer}. This assumption translates into freedom for variables and parameters describing the monomers to take on any value, in our mathematical model below. The above premise statement has its roots in the supervenience of evolution over chemistry. Although RNA is widely thought to have evolutionarily preceded DNA and is thus better situated for evolution-based explanations, we are constrained to concentrate on DNA, due to the comparative lack of experimental information on the thermodynamics and kinetics of non-enzymatic RNA double-strand formation/unzipping\cite{RNAreplicationSzostak}.

\section*{The model}
In our simple phenomenological model of a primordial self-replicating system (Methods), we consider an autocatalytic polymer that is capable of replicating without the help of enzymes. A single strand of the polymer catalyzes the formation of another strand on top of itself, by functioning as the template. Free-floating monomers attach to the bound monomers on the template strand at lower temperatures, and facilitate covalent bonding between monomers\cite{andersonTempCycling} and hence polymerization, leading to the formation of the replica strand. The replica strand dissociates from the template strand at higher temperatures, creating two single strands, as happens in a Polymerase Chain Reaction. 

A self-replicating molecular species must satisfy certain requirements in order to be evolutionarily successful and to function as an information-carrier. In the following, we list those physically meaningful requirements to be satisfied by the molecular species, and in doing so, arrive at two conflicting requirements. Breaking of a symmetry, upon maximization of replicative potential, leads to resolution of the conflict and to simultaneous satisfaction of the two requirements. These requirements are not new, and have been included and explored individually in other models and systems elsewhere\cite{andersonTempCycling,model_like_mine,similartomine,Toymagneticmodel}. 


Self-replication involves both bond formation between free-floating monomers and monomers on the template strand, and bond-breaking between monomers on the two strands, requiring these inter-strand bonds to be relatively weak compared to other bonds in the polymer. On the other hand, information storage requires stronger intra-strand bonds that withstand strong environmental variations, as pointed out by Schr{\"o}dinger\cite{whatislife}. Hence, the self-replicating polymer needs to be composed of two complementary components, mutable inter-strand ``hydrogen bonds'' and relatively immutable intra-strand ``covalent bonds''\cite{andersonTempCycling,model_like_mine,similartomine,Toymagneticmodel}.

The intrinsic covalent bonding rates among free-floating monomers should be lower than the covalent bonding rates between the monomers hydrogen-bonded to the template strand, so that monomers become available for self-replication and not for \textit{de novo} strand formation. This requirement makes self-replication viable and information transfer across generations possible. Evolution could have solved this by identifying monomers whose kinetic barrier for covalent bonding between themselves is lowered when they are attached to the template strand\cite{andersonTempCycling,weak_hydrogenbonds,model_like_mine}. We term this barrier reduction ``\textit{covalent bond catalysis}''.

If a hydrogen bond catalyzed the formation (and hence dissociation as well) of another hydrogen bond in its neighborhood\cite{tidalcycling2}, the strand would be replicatively more successful, since covalent bond formation requires two contiguous monomers hydrogen-bonded to the template. Also, higher rate of monomer attachment to the template would allow for more monomers to be drawn in for polymerization, away from other competing processes such as dimerization through hydrogen bonding. Thus, reduction of kinetic barrier for hydrogen bond formation would be advantageous for the self-replicating system. The foregoing justifies the need for ``\textit{hydrogen bond cooperativity}'', catalysis of hydrogen bond formation/dissociation by their neighboring hydrogen bonds\cite{anticooperative_bonding,polandsheragamodel,PBDmodel,basepairingphysics}. Aforementioned cooperativity, the increasing ease of hydrogen bonding between unbonded monomers (zippering) when two single strands are already hydrogen-bonded at one of the ends, is a very well-established phenomenon in DNA, and has been well-studied both experimentally and theoretically\cite{basepairingphysics}. The experimental signature of cooperativity in DNA melting is the sharpness of the melting transition, where the DNA goes from a double strand to two single strands within a narrow range of temperature\cite{sharpmeltingdna}. Cooperativity in DNA has also been abundantly documented in DNA zipping and unzipping experiments\cite{sequencedependence1999,cooperativePNAS,cooperativitypnas2,cooperativityPRL}. The presence of cooperativity in RNA double-strand is an open question due to the lack of such unzipping experiments on double-stranded RNA, to our knowledge.

Obviously, the probability for the covalent bond formation between two contiguous monomers on the replica strand will increase with the lifetime of the hydrogen bonds of the monomers with the template strand. Thus, higher the kinetic barrier for hydrogen bond dissociation, higher the probability for the successful formation of the covalent bond and hence the replica strand. Thus, we notice that, \textit{while covalent bond catalysis requires higher kinetic barrier for hydrogen bond dissociation, hydrogen bond cooperativity requires lower kinetic barrier for hydrogen bond formation}. Since self-replication requires the replicating polymer to be at or near the melting point of the hydrogen bonds, the kinetic barriers for formation and dissociation are nearly equal, and we arrive at the competing requirement of both higher and lower kinetic barrier height, or equivalently, to fine-tuning of the hydrogen bond lifetime. We could solve this conundrum by introducing an environment with oscillating ambient temperature, where, the hydrogen bond lifetime is longer at lower temperatures and thus enables covalent bond formation, whereas, higher temperatures facilitate strand separation. Nevertheless, strands that \textit{intrinsically} satisfy these two competing requirements would still be evolutionarily more successful, by being able to colonize regions with temperature oscillations of much smaller amplitude.

The solution that simultaneously and intrinsically satisfies these two competing requirements is to break the symmetry\cite{moreisdifferent} of the catalytic influence of a hydrogen-bonded monomer-pair on its two neighboring hydrogen bonds on either side. The hydrogen-bonded monomer-pair can reduce the kinetic barrier for hydrogen bond formation/dissociation to its right, while increasing the barrier for hydrogen bond formation/dissociation to its left, (or vice versa) which we call ``asymmetric hydrogen bond cooperativity''. This solution is similar in spirit to Kittel's single-ended zipper model for DNA\cite{Kittelsmodel}. Asymmetric cooperativity has also been proposed earlier to explain other biophysical processes\cite{asymmetriccooperativity}. Such an arrangement would prolong the lifetimes of the already-formed hydrogen bonds to the pair's left, and thus would increase the probability for covalent bonding among those bonded monomers. It will also enable rapid extension of the replica strand to the right, drawing monomers away from competing processes, by allowing monomers to hydrogen bond with the template easily through the reduction of the kinetic barrier. Thus, the broken symmetry of unequal and non-reciprocal catalytic influence leads to simultaneous satisfaction of the above-mentioned two competing requirements. Surprisingly, the replicative advantage of strands with asymmetric cooperativity over symmetric strands turns out to be crucial for understanding various intrinsic physico-chemical properties of the extant heteropolymer, DNA. Again, due to the lack of information about the mechanisms of RNA double-strand construction and unzipping, we will not have much to say about the evolutionarily earlier self-replicating heteropolymer, RNA, and have to confine oursleves to DNA.

Our model (methods) simply translates the foregoing in mathematical language. We imagine the construction of a replica strand of an autocatalytic polymer on top of the template strand as a Markov Chain. A Markov chain description of a random process involves identification of the state space, and writing down the transition rates or probabilities between the identified states. Given the transition rate matrix, we can calculate variables that are relevant for our analysis, such as the average first passage time to a given state and average residence time in a given state (methods). We measure the potential of a molecular species to form a replica strand as the product of two factors: the relative rate of monomer utilization for replica strand formation against other competing processes, and the probability for covalent bond formation between any two monomers on the replica strand. The first factor increases with reduction in hydrogen-bonding kinetic barrier, whereas the second factor decreases with the reduction in the barrier height. Asymmetric cooperativity simultaneously satisfies both the requirements, as we show in the next section.

It is crucial to understand that our goal for building this model is limited to demonstrating, with minimal assumptions and in a physically transparent manner, the superiority of primordial self-replicating polymers with asymmetric cooperativity over polymers with symmetric cooperativity in attracting enough monomers to construct the replica strand. In particular, we do not intend for this model to make quantitative predictions about the kinetics of DNA (un)zipping or helix-coil transition, for which, highly sophisticated models already exist\cite{dnastructurefunction}. In keeping with this limited goal, we have included in the model only ingredients that have a direct bearing on our aforementioned goal. We exclude all other ingredients that provide negligible or no discriminatory capability, even though they might make the model more realistic and accurately reflective of the self-replication process. The ingredients that we reasoned to have the same effect on the self-replication of both symmetric and asymmetric polymers, and is thus non-discriminatory, such as polymer bending, secondary structure formation, multiple monomer types, inclusion of N-mers and so on were thus excluded. In particular, we ignore the differences in the rates of unzipping between the symmetric and asymmetric-cooperative versions of the double strand, during the high temperature phase of the temperature cycle, by assuming that the time period of the temperature cycles are much larger than the zipping and unzipping times of the double strand, in order to keep the model as simple as possible. This assumption minimizes the contribution of unzipping rates to replicative potential, allowing us to solely concentrate on the competition between symmetric and asymmetric polymers for monomer precursors. It can be argued that the rate of self-replication of a given polymer species is also determined by other processes in its self-replication cycle, such as the formation rates of its monomers from their precursors, the cleavage of the monomers and polymers, attachment of monomers on wrong templates and so on. In the absence of any rationale for faster production of symmetric monomers over asymmetric monomers from common precursors, the variation in the above-mentioned rates of other legs of the self-replication cycle is similar for both symmetric and asymmetric monomer classes, which is non-discriminatory and hence can likewise be ignored.

\section*{Methods}
Our aim here is to encapsulate in mathematical language the sequence of hydrogen and covalent bonding and unbinding events that result in non-enzymatic self-replication of the autocatalytic polymer. We assume a circular or linear template polymer, constructed by stringing together $N$ monomers through covalent bonding. Free-floating monomers can either hydrogen-bond with each other, forming dimers, at a rate $r_f$, or can bind with the template to initiate the construction of the replica strand, at a rate $r_{t0}$. We denote the presence or absence of a hydrogen bond between a monomer in the replica strand and the $i$-th monomer on the template strand with a $1$ or $0$ in the $i$-th place in a binary string of $N$ digits. Thus, for $N=5$, the binary string $00000$ would imply that the template strand has no monomers hydrogen-bonded to it, and $00100$ implies one monomer hydrogen-bonded to the third monomer on the template strand. Cooperativity of hydrogen bonding is implemented by stipulating different rates for subsequent monomer binding events, depending upon the presence or absence of neighboring hydrogen bonds. The rates $\mathcal{R}$, of monomers hydrogen bonding with template strand in different hydrogen-bonding neighborhoods can then be expressed as
\begin{equation}
\begin{aligned}
& \mathcal{R}\left ( 00000 \rightarrow 00100 \right ) = r_{t0}, \\
& \mathcal{R}\left ( 00100 \rightarrow 00110 \right ) = r_{tr} = \alpha_R r_{t0}, \\
& \mathcal{R}\left ( 00100 \rightarrow 01100 \right ) = r_{tl} = \alpha_L r_{t0} \qquad \text{and}\\
& \mathcal{R}\left ( 01010 \rightarrow 01110 \right ) = r_{tc} = \alpha_R \alpha_L r_{t0}. \\
\label{ratesbonding}
\end{aligned}
\end{equation}

The unbinding rates are

\begin{equation}
\begin{aligned}
& \mathcal{R}\left ( 00000 \leftarrow 00100 \right ) = s_{t0}, \\
& \mathcal{R}\left ( 00100 \leftarrow 00110 \right ) = s_{tr} = \alpha_R s_{t0}, \\
& \mathcal{R}\left ( 00100 \leftarrow 01100 \right ) = s_{tl} = \alpha_L s_{t0} \qquad \text{and}\\
& \mathcal{R}\left ( 01010 \leftarrow 01110 \right ) = s_{tc} = \alpha_R \alpha_L s_{t0}. \\
\label{ratesunbinding}
\end{aligned}
\end{equation}
In Eqs. \ref{ratesbonding} and \ref{ratesunbinding}, $\alpha_R$ and $\alpha_L$ are the factors that modify the rates of hydrogen bonds forming to the right and left of a single hydrogen bond. Symmetric cooperativity results when $\alpha_L=\alpha_R$, and when these two factors are unequal, asymmetric cooperativity results. If we assume that only nearest neighbor hydrogen bonds affect the rate of bonding of another monomer to the template strand, the above rates of bond formation and dissociation are sufficient to determine the rates of transition between all $2^5=32$ states that describe the $N=5$ double-strand formation process.

The rate constants for the transition between all possible states describing the double-strand formation are determined by just four parameters $r_{t0}$, $s_{t0}$, $\alpha_R$, and $\alpha_L$. We analyze part of the self-replication process as a continuous-time Markov Chain process. We can evaluate the average time it would take for the template strand to go from one without any monomers attached to it, to one with all of its monomers hydrogen-bonded to a monomer, i.e., from the state $00000$ to $11111$. This is calculated using the well-established ``first passage time'' or ``hitting time'' analysis\cite{stochasticprocesses}: Let $\mathcal{R}_{ij}$ be the transition rate constant from state $i$ to $j$, and $t_i$ the average time taken for the Markov chain to reach the final state $k=11111$ when it begins at state $i$. Then, first passage time analysis involves solving the following set of linear equations for non-negative $t_i$'s:
\begin{equation}
\sum_j \mathcal{R}_{ij} (t_j - t_i) = -1, \qquad i \ne k.
\label{hittingtimeeq}
\end{equation} 
The average time taken to traverse from $00000$ to $11111$, $t_1$ in Eq. \ref{hittingtimeeq}, is henceforth called the ``growth time'' $t_g$. The ``rate advantage'', a measure of the propensity for monomers to hydrogen-bond with a template as opposed to hydrogen-bonding among themselves, is 
\begin{equation}
P_g = \frac{1/t_g}{1/t_g + r_f},
\label{hittingtime}
\end{equation}
where, $r_f$ is the rate of dimerization of monomers.

Let the rate of covalent bond formation between two contiguous monomers attached to the template strand be $r_c$. The probability for the covalent bond to form within a certain time $t$ is then
\begin{equation}
P_c = 1-exp \left ( - r_c t\right ).
\label{probcov}
\end{equation}
The average lifetime of the configuration of a pair of contiguous monomers hydrogen-bonded to the template strand determines the probability of a covalent bond forming between the two monomers, through the above Eq. \ref{probcov}. The most conservative estimate of such a lifetime is the lifetime of the state $11111$, because the last covalent bond has the least time to form, since all other pairs of monomers have been in existence before the last pair. The average lifetime of the state $11111$ is just $1/\mathcal{R}_{11111}$, the inverse of the diagonal entry corresponding to the state $11111$ in the transition rate matrix. The expression for $P_c$ then becomes
\begin{equation}
P_c = 1-exp \left ( - \frac{r_c}{\mathcal{R}_{11111}}\right ).
\label{probcov2}
\end{equation}
  
	While low barrier height for bonding decreases the ``first passage time'' $t_g$ and thus increases the rate advantage $P_g$, it would decrease the covalent bonding time $P_c$. Both fast ``growth time'' (measured by $P_g$) and successful covalent bonding (measured by $P_c$) are important for the success of a self-replicating polymer in creating a full replica strand, which can be measured using the dimensionless metric $P = P_g P_c$, called ``replicative potential'' in this paper. But the conflicting requirements for both these metrics to maximize their respective values, with $P_g$ maximization requiring $\alpha_L, \alpha_R > 1$, and $P_c$ maximization requiring $\alpha_L, \alpha_R <1$ , sets up a conflict. The conflict is resolved when the left-right symmetry is broken upon maximization of the replicative potential, with $\alpha_L <  \alpha_R$ or $\alpha_L > \alpha_R $.  
	
	\subsubsection*{Parametrization}
	The model involves two species-specific timescales: $r_{t0}(s)$ and $s_{t0}(s)$, the average rates for formation and dissociation of uncatalyzed hydrogen bonds (both of which are of the same order, given the fact that self-replication is maximally successful in temperature cycles that include the melting point of the hydrogen bonds), and $r_c(s)$, the average rate of formation of a covalent bond between two contiguous monomers hydrogen bonded to the template strand, of a polymer species $s$.  This suggests that we examine three distinct parameter regimes: (i) $r_{t0}(s) > r_c(s)$, (ii) $r_{t0}(s) \approx r_c(s)$ and (iii) $r_{t0}(s) < r_c(s)$. Since the rate of formation of a hydrogen bond between a free-floating monomer and a monomer attached to the template, $r_{t0}(s)$, depends on the monomer concentration in the primordial soup, the above three regimes can all be reached by varying the monomer concentration. The \textit{non-enzymatic} rate constant for the above hydrogen bonding has been measured to be of the order of $10^9 M^{-1} min^{-1}$ for DNA, at pH $7$ and $279 K$\cite{hydrogenbonding2}. The \textit{non-enzymatic} rate of extension of the replica strand, through covalent bonding between two activated nucleotides hydrogen-bonded to the template, $r_c$, has been measured to be of the order of $10^{-2} min^{-1}$ at pH $8.9$ and $293 K$\cite{incorporationkinetics}. During self-replication, as the concentration of templates of different polymer species increased exponentially/supralinearly, the competition between symmetric and asymmetric polymer species for monomer precursors and energetic sources would have intensified at progressively lower respective monomer concentrations. We choose low values for monomer concentrations, where this competition is operative. The hydrogen bonding rates $r_{t0}$, at $1 nM$, $0.01 nM$ and $1 pM$ concentrations, are then evaulated to be of the order of $1 min^{-1}$, $10^{-2} min^{-1}$ and $10^{-3} min^{-1}$ respectively. These three rates correspond to the three distinct parameter regimes mentioned above.

		Since our goal is to just demonstrate the replicative superiority of asymmetric polymers, and not quantitative predictions, we ignore the difference in the types of nucleotides used and the values of the environmental variables (pH and temperature) between the above two experiments. In any case, the model is relatively insensitive to the precise values used for the parameters. For simplicity, the rate of dimerization through hydrogen-bonding of two monomers, $r_f$, is taken to be the same as the rate of monomer attaching to the polymer template, $r_{t0}$. The dimensionless catalytic/inhibitory factors $\alpha_L$ and $\alpha_R$ are allowed to independently vary between $0.1$ and $4$\cite{incorporationkinetics}, allowing us to continually interpolate between the symmetric Ising-type interactions and the asymmetric Zipper-type interactions. We choose the monomer unbinding rate $s_{t0}$ = $r_{t0}/2$, and thus implicitly assume that the replicating polymer is at a temperature slightly below the melting point of the hydrogen bonds, conducive for replica strand construction.

		

\section*{Results}

\begin{figure}
\begin{center}
\includegraphics[width=0.9\textwidth]{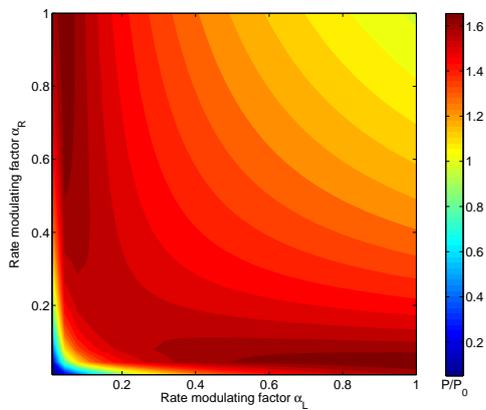}
\caption{Replicative potential $P$ of circular self-replicating polymer strands of length $N=5$, as a function of factors $\alpha_L$ and $\alpha_R$ modulating the rates of hydrogen bonding/unbinding to the left and right of a single pre-existing hydrogen bond. The parameter regime for this plot is characterized by hydrogen bonding rates higher than covalent bonding rates, $r_{t0} > r_c$. Maximum replicative potential is achieved when the bonding rate of both the left and right bonds are reduced, albeit unequally, in order to increase the covalent bond formation probability without significantly impacting the monomer bonding rate. This results in the breaking of left-right symmetry. The two equivalent maxima in the figure, where $\alpha_L \ne \alpha_R$ and $\alpha_L, \alpha_R < 1$, correspond to two equally possible modes of asymmetric cooperativity. This symmetry-breaking is the consequence of a compromise between two competing requirements for successful self-replication: rapid hydrogen-bonding and unbinding of monomers with the template to speed up replication, and successful formation of covalent bonds between two contiguous hydrogen-bonded monomers, which depend on long hydrogen-bond lifetimes. The replicative potential $P$ above is measured in units of $P_0= P(\alpha_L=1, \alpha_R=1)$, the replicative potential without hydrogen-bond cooperativity.}
\label{regime1}
\end{center}
\end{figure}

\begin{figure}
\begin{center}
\includegraphics[width=0.9\textwidth]{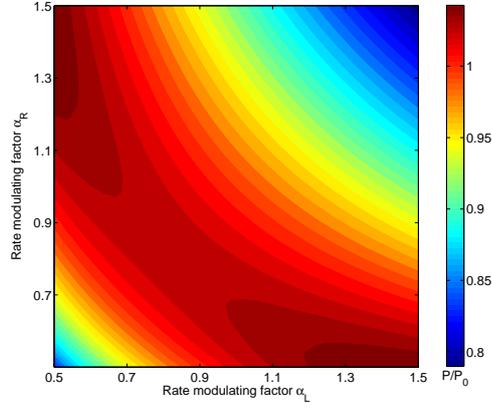}
\caption{Replicative potential $P$ of circular self-replicating polymer strands, in the case of nearly equal hydrogen bonding and covalent bonding rates, $r_{t0} \approx r_c$. Maximum replicative potential is achieved when the bonding rate of the left bond is reduced as much as possible, and the bonding rate of the right bond is increased above the uncatalyzed rate (or vice versa), resulting in the breaking of left-right symmetry. Species with maximal replicative potential  are located at $\alpha_L > 1$ and $\alpha_R < 1$ (or vice versa) in the figure. The two factors that go into calculation of $P$, the rate advantage $P_g$ and the covalent bonding probability $P_c$, are shown in figs. \ref{regime2_Hbonding} and \ref{regime2_Cbonding}.}
\label{regime2}
\end{center}
\end{figure}

\begin{figure}
\begin{center}
\includegraphics[width=0.9\textwidth]{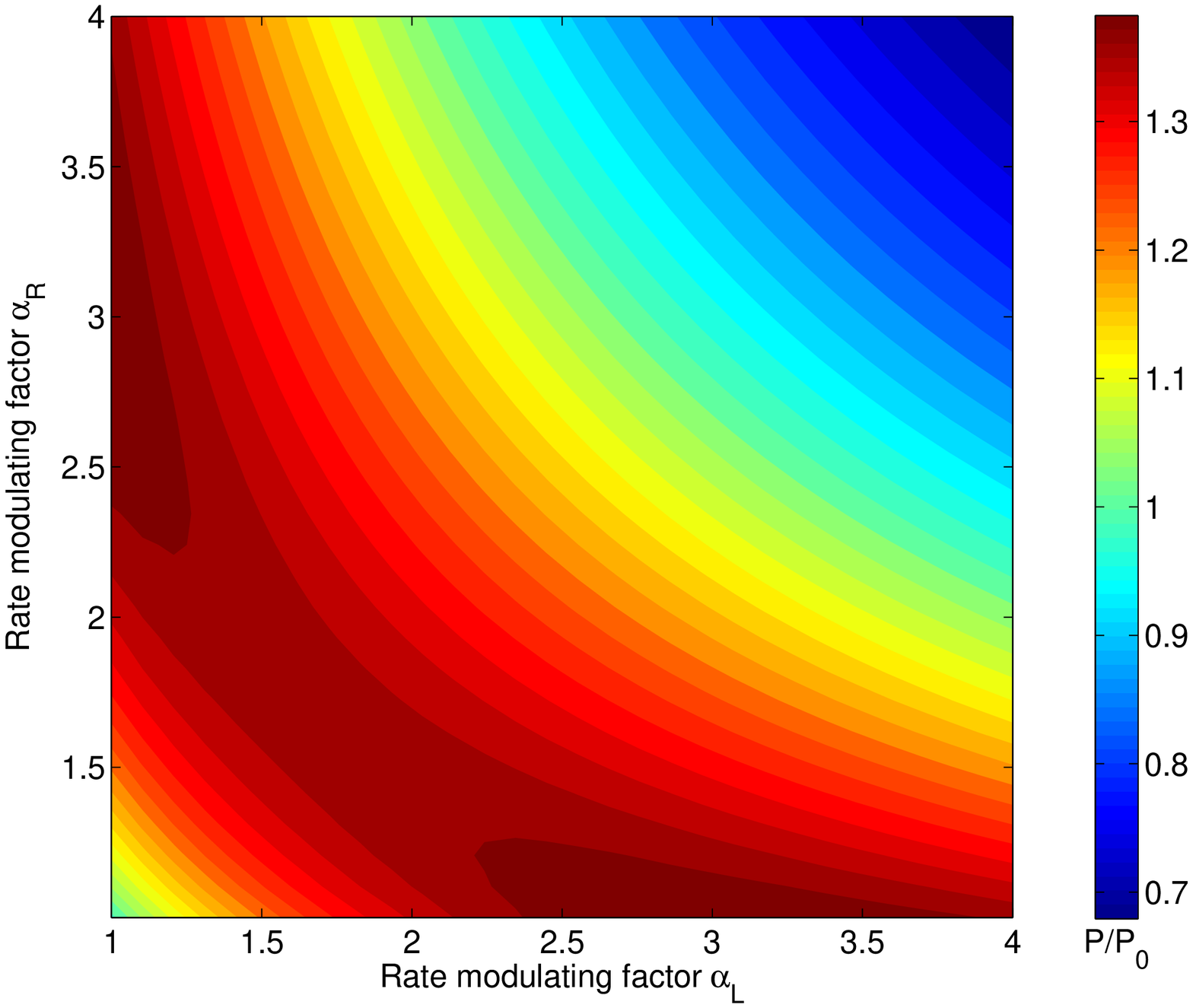}
\caption{Replicative potential $P$ of circular self-replicating polymer strands, in the regime where covalent bonding rate is higher than hydrogen bonding rate, $r_{t0} < r_c$. Maximum replicative potential is achieved when the kinetic barriers of both left and right bonds are decreased, albeit unequally, in order to simultaneously increase the strand growth rate and covalent bonding probability. Thus the maxima are located at $\alpha_L, \alpha_R < 1$ and $\alpha_L \ne \alpha_R$.}
\label{regime3}
\end{center}
\end{figure}

In this section, we show that the competing requirements mentioned above for evolutionary success in self-replication of circular polymers lead to breaking of the symmetry of catalytic influence of a hydrogen bond on its neighbors on either side, in all the three parameter regimes mentioned above. Figures \ref{regime1}, \ref{regime2} and \ref{regime3} show the normalized replicative potential $P$ as a function of two variables $\alpha_L$ and $\alpha_R$, the catalytic/inhibitory factors modulating the bonding rates of hydrogen bonds to the left and right of a single pre-existing hydrogen bond, in the three parameter regimes mentioned above. A point in the plot, say fig. \ref{regime1}, represent a specific species with its own catalytic/inhibitory factors $\alpha_L$ and $\alpha_R$. The species' replicative potential is represented by the color value at that point. All the three plots show two maxima, both equally off the diagonal where the bonding rates are equal, proving our assertion that species with asymmetric cooperativity are replicatively more successful. This is a \textit{genuine} symmetry-breaking, since \textit{two equivalent degenerate maxima} are present on either side of the symmetric cases (along the diagonal from lower left to top right in figs. \ref{regime1}, \ref{regime2} and \ref{regime3}), and both solutions are equally probable. The role played by energy minimization in symmetry-breaking in non-living systems is played here by evolution, i.e., replicative potential maximization. This symmetry-broken solution is quite insensitive to the values of parameters used, as long as the conflict of requirement for both high and low kinetic barriers remain in effect.

	In the first parameter regime , where $r_{t0}(s) > r_c(s)$, the rate constant for covalent-bond formation is lower than that of uncatalyzed hydrogen bonding/unbinding. The hydrogen-bonded monomers need to stay longer on the template to allow for covalent bond to form, which implies that the need for longer hydrogen bond lifetime is stronger than the need for monomer acquisition, for successful self-replication. Thus, successul polymers inhabiting this regime will have hydrogen bonds whose kinetic barrier for formation/dissociation is increased by neighbors from both sides, albeit unequally, in a compromise between the conflicting needs for both higher rates of hydrogen bonding/unbinding and higher covalent bond formation probability. This implies that $\alpha_L, \alpha_R < 1$ and $\alpha_L \ne \alpha_R$ in this regime, as shown in Fig. \ref{regime1}.

	In the second regime, where $r_{t0}(s) \approx r_c(s)$, the rates of uncatalyzed hydrogen bond formation/dissociation and covalent bond formation are nearly equal, which implies that the requirements for both higher rates of hydrogen bonding/unbinding and higher covalent bond formation probability are nearly equally important. Symmetry breaking apportions the rate-modifying factors $\alpha_L$ and $\alpha_R$ appropriately between the two requirements above, with $\alpha_L>1$ and $\alpha_R<1$ or vice versa, as shown in Fig. \ref{regime2}. Higher kinetic barrier for hydrogen bond formation to the left of a pre-existing hydrogen bond results in hydrogen bonds with longer lifetimes and hence higher covalent bond formation probability, whereas, lower barrier to the right enables monomers to attach easily to the template, resulting in faster elongation of the replica strand.

	In the third regime, where $r_{t0}(s) < r_c(s)$, the rate of covalent bond formation is higher than the rate of hydrogen bond formation/dissociation. Thus, the template strand's ability to acquire monomers for replica strand growth can be increased further, without significantly impacting the covalent bond formation rates. This is done by reducing the kinetic barrier for hydrogen bond formation/dissociation from both left and right neighboring bonds, while keeping the two rates unequal to increase covalent bonding probability. This implies that $\alpha_L, \alpha_R > 1$ and $\alpha_L \ne \alpha_R$ in this regime, as shown in Fig. \ref{regime3}.

\begin{figure}
\begin{center}
\includegraphics[width=0.9\textwidth]{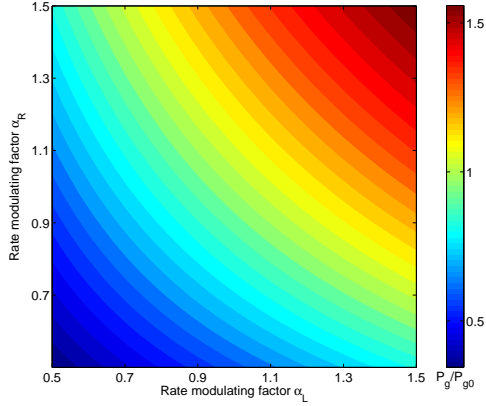}
\caption{Rate advantage $P_g$, normalized with respect to that of a strand with no hydrogen bond cooperativity, as a function of rate-modulating factors $\alpha_L$ and $\alpha_R$, in the parameter regime $r_{t0} \approx r_c$. More monomers can be drawn in for template-directed polymerization if the rate of hydrogen-bonding of monomers with the template is high. The maximum rate advantage occurs at the highest possible values of the bonding rates $r_{tl}$ and $r_{tr}$, and where $r_{tl}=r_{tr}$ or $\alpha_L=\alpha_R$.}
\label{regime2_Hbonding}
\end{center}
\end{figure}

\begin{figure}
\begin{center}
\includegraphics[width=0.9\textwidth]{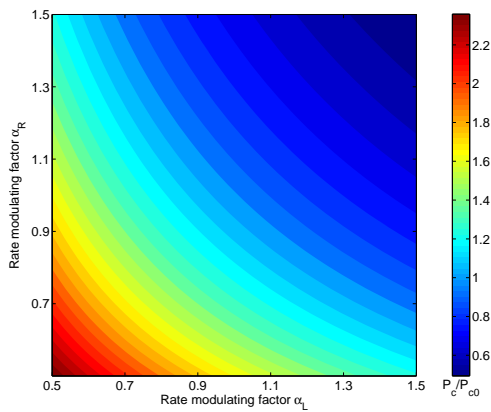}
\caption{Covalent bonding probability $P_c$, normalized with respect to that of a strand with no hydrogen bond cooperativity, as a function of rate-modulating factors $\alpha_L$ and $\alpha_R$, in the regime $r_{t0} \approx r_c$. Long lifetime of a pair of contiguous hydrogen bonds increase the covalent bonding probability between the two monomers. High $P_c$ requires low unbinding rates and hence high kinetic barriers near hydrogen-bond melting point. Maximum of covalent bonding probability occurs at the lowest possible values of the bonding rates $r_{tl}$ and $r_{tr}$, and where $r_{tl}=r_{tr}$ or $\alpha_L=\alpha_R$.}
\label{regime2_Cbonding}
\end{center}
\end{figure}

As we mentioned earlier, the replicative potential $P$ is the product of two conflicting factors:  (1) The rate of monomer utilization for polymerization relative to the combined rate of all processes requiring the monomers $P_g$. This relative rate depends upon the rate of monomers hydrogen bonding with monomers on the template strand. The lower the effective barrier height for hydrogen bonding, higher will be the rate of monomer utilization. This is illustrated in Fig.  \ref{regime2_Hbonding}, which shows that higher bonding rates of the left and right neighbors lead to higher utilization, and maximum utilization occurs when both left and right rates are \textit{equal}. The maxima are located at points where $\alpha_L = \alpha_R$ and $\alpha_L, \alpha_R > 1$. (2) The probability for covalent bonding $P_c$. This depends on the average lifetime of two contiguous hydrogen bonds, and, higher the barrier height for hydrogen bond dissociation, higher the probability for covalent bonding. This is illustrated in Fig.  \ref{regime2_Cbonding}, where the probability is seen to be high for lower rates of hydrogen unbinding, and when both the left and right unbinding rates are \textit{equal}. Since self-replication only happens near hydrogen bond melting point, the bonding and unbinding rates are of the same order. The maxima are located at points where $\alpha_L = \alpha_R$ and $\alpha_L, \alpha_R <1$. Figs. \ref{regime2_Hbonding} and \ref{regime2_Cbonding} show that the two factors $P_g$ and $P_c$, whose product is the replicative potential $P$, cannot be simutaneously maximized, since they conflictingly require high and low hydrogen bonding/unbinding rates, for their respective maximizations. The replicative potential maxima instead happen where the bonding rates of the hydrogen bonds to the left and right of a single pre-existing hydrogen bond are \textit{unequal}, with $\alpha_L \ne \alpha_R$ and $\alpha_L > 1, \alpha_R < 1$ or vice versa. This broken symmetry solution provides explanations for multiple fundamental properties of DNA, as we describe below.

\begin{figure}
\begin{center}
\includegraphics[width=0.9\textwidth]{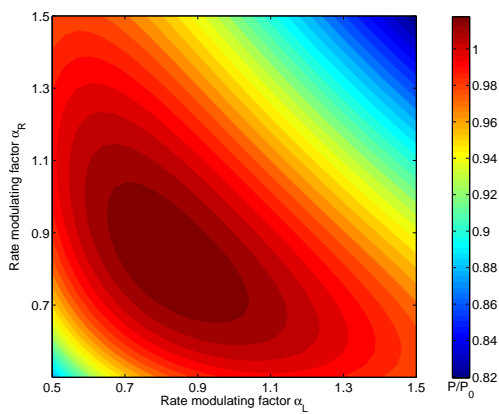}
\caption{Replicative potential $P$ of linear self-replicating polymer strands as a function of factors $\alpha_L$ and $\alpha_R$ modulating the rates of hydrogen bonding to the left and right of a single pre-existing hydrogen bond. Maximum replicative potential occurs when the bonding rates of left and right bonds are equal, $r_{tl}=r_{tr}$ or $\alpha_L=\alpha_R$. This maximum value is lower than the maximum values in the circular strand case, demonstrating the replicative superiority of circular strands over linear strands. The replicative potential $P$ above is measured in units of $P_0= P(\alpha_L=1, \alpha_R=1)$. Broken symmetry compromise for the two competing requirements of self-replication is unviable in linear strands due to the high kinetic barrier in one direction (needed for covalent bond formation) inhibiting hydrogen bonding and thus preventing strand growth in that direction.}
\label{regime2_lin}
\end{center}
\end{figure}

Interestingly, only circular strands adopt the above broken-symmetry solution, to satisfy the two competing requirements. The maximal replicative potential of linear strands is lower than the cicular strands' maxima, and occurs where the bonding rates of left and right hydrogen bonds are equal, i.e., when the strands are symmetrically cooperative with $\alpha_L = \alpha_R$, as shown in Fig.  \ref{regime2_lin}. The reason behind the difference between circular and linear strand behavior is as follows. In the circular strand case, the first hydrogen bond connecting a free-floating monomer and a monomer on the template strand can form at any monomer position on the strand. Whereas, in the linear strand case, the first hydrogen bond must form at the rightmost end, if the strand is to self-replicate as effectively as the circular strand. Since asymmetric cooperativity increases the barrier for hydrogen bond formation to the right, formation of the first hydrogen bond at any location other than the right-most template monomer will result in severe inhibition of bond formation to the right of that first bond (replacing ``right'' with ``left'' results in an equally valid statement). This reduces the effectiveness of self-replication, and thus disincentivizes the adoption of asymmetric cooperativity as a solution for satisfying the two competing requirements, in linear strands. 

\clearpage

\section*{Experimental support for asymmetric cooperativity}
The \textit{central prediction} of our model above is the presence of asymmetric cooperativity in evolutionarily successful self-replicating polymers, which includes DNA. Asymmetric cooperativity, unequal catalysis of hydrogen bonds on the left and right, can manifest itself  by \textit{rendering kinetics of zipping of two DNA single-strands and unzipping of DNA double-strands from the left and the right ends, unequal}. This is an eminently experimentally observable phenomenon.

We first point to the evidence for the presence of \textit{directional asymmetry} , the inequivalence of left and right side in DNA, because of it being a well-established fact in Biology. We would like to clarify that the term ``directional asymmetry" is not equivalent to ``asymmetric cooperativity". The former is more generic and can be of thermodynamic and/or kinetic origin, whereas the latter is purely of kinetic origin. In this paragraph, the evidence provided is for the generic directional asymmetry, and we defer differentiating between the asymmetry's thermodynamic and kinetic origins to the following paragraphs. Let us denote the base-pairing of nucleotides (the four different types of monomers present in DNA) on the top and the bottom strands of the double-stranded DNA as $5' \hyp X \hyp 3'/3' \hyp Y \hyp 5'$, with $5' \hyp X \hyp 3'$ in the top strand hydrogen-bonded to $3' \hyp Y \hyp 5'$ in the bottom strand. The growth of a replica strand on a single strand template DNA happens only in one direction, and the numbers $5'$ and $3'$ are used to identify that direction. Directional asymmetry in DNA can be easily demonstrated by using the well-established nearest-neighbor \textit{thermodynamic} parameters of DNA\cite{ThermodynamicAsymmetry}, wherein, the free energy, enthalpy and entropy of different combinations of nearest-neighbor pairs were experimentally measured and cross-verified. It can be seen from the tables in \cite{ThermodynamicAsymmetry} that, adding $5' \hyp G \hyp 3'/3' \hyp C \hyp 5'$ base-pair to the \textit{left} of  $5' \hyp A \hyp 3'/3' \hyp T \hyp 5'$, resulting in $5' \hyp GA \hyp 3'/3' \hyp CT \hyp 5'$, and adding the same $5' \hyp G \hyp 3'/3' \hyp C \hyp 5'$ to the \textit{right} of $5' \hyp A \hyp 3'/3' \hyp T \hyp 5'$, resulting in $5' \hyp AG \hyp 3'/3' \hyp TC \hyp 5'$, are different operations, result in distinct chemical structures, and obviously have different nearest-neighbor \textit{thermodynamic} parameters. Thus our asymmetric cooperativity prediction merely extends such asymmetric thermodynamic influence to \textit{kinetics}. A note on terminology: The inter-strand bonding between the nucleotides $A$ and $T$ is composed of two hydrogen bonds, and between $G$ and $C$, three hydrogen bonds. Since we have no need to distinguish between either of the two hydrogen bonds between $A$ and $T$ or between the three bonds between $G$ and $C$, we collectively refer the bonds between $A$ and $T$, and between $G$ and $C$ in singular, as ``a hydrogen bond''. Thus, ``interactions between neighboring hydrogen bonds'' would imply interaction between hydrogen bonds of two neighboring base-pairs, and not between the hydrogen bonds of a single base-pair.

\begin{figure}[htbp]
\hspace*{-2cm} 
\includegraphics[width=1.2\textwidth]{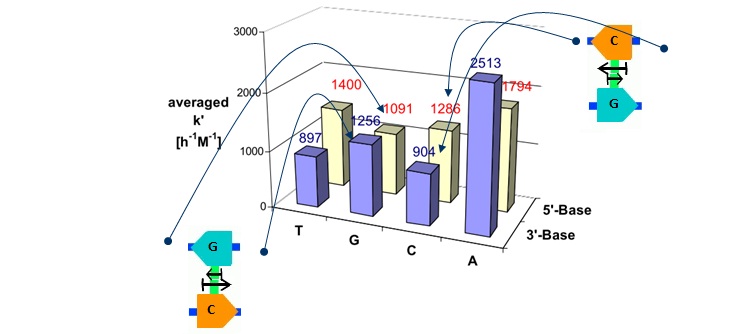}
\caption{The bar plot, reproduced here with permission from \cite{incorporationkinetics}, shows the experimentally observed dependence of rate of extension of a template-attached primer by a single nucleotide on its neighboring nucleotides. The rate of extension is higher when the nucleotide $G$ is the immediate neighbor on the $3'$ side of the primer, or when $C$ is the $5'$ side immediate neighbor on the downstream binding nucleotide. With $G$ at the $5'$ side and $C$ at the $3'$ side, the rates are lower. In other words, the base-pair $5' \hyp C \hyp 3'/3' \hyp G \hyp 5'$ supports higher rate of nuceotide incorporation to its left compared to $5' \hyp G \hyp 3'/3' \hyp C \hyp 5'$, whereas $5' \hyp G \hyp 3'/3' \hyp C \hyp 5'$ supports higher incorporation rate to its right compared to $5' \hyp C \hyp 3'/3' \hyp G \hyp 5'$. Thus the orientation of the base-pair dictates the direction of asymmetric cooperativity, and the latter agrees with the direction of catalysis and inhibition derived from the relationship between replication orientation and $GC$ skew. The asymmetric cooperativity also exists in $AT$ base-pair, but with an added constant component when the nucleotide A is present on the replica strand neighborhood.}
\label{kerviokinetics}
\end{figure}

A crucial piece of evidence for the existence of directional asymmetry in the \textit{kinetics} of DNA, i.e., asymmetric cooperativity, comes from studying the incorporation kinetics of activated nucleotides that \textit{nonenzymatically} extend a primer attached to a template strand, one nucleotide at a time, in the presence of a downstream binding strand\cite{incorporationkinetics}. It is convincingly shown here by Kervio et al that the kinetics of the extension of a primer on top of a template strand (which include hydrogen and covalent bond formation) depends on the local sequence, in a way that strongly corroborates our hypothesis of the presence of asymmetric cooperativity in DNA. Fig. S6 in their paper, reproduced here as Fig. \ref{kerviokinetics}, clearly shows the asymmetry in the kinetics.  

More experimental evidence for asymmetric cooperativity come in part from unzipping experiments. In one experiment\cite{lambdadnaevidence}, a single-molecule phage $\lambda$ DNA is unzipped using force applied on a microscopic glass slide attached to it. The measured forces of unzipping from one end is shown to be different from unzipping from the opposite end, and this is explained by the group as due to the presence of stick-slip motion\cite{lambdadnaevidencefull}. The fact that different forces signatures are needed to unzip the DNA molecule from either end implies that the work done to unzip the DNA from either end is also different. Under the near-equilibrium experimental conditions of unzipping as mentioned in the article\cite{lambdadnaevidence}, this difference in the unzipping forces cannot be due to thermodynamics. Thus the difference can only be due to the difference in kinetics of unzipping from either end, strongly supporting the presence of asymmetric cooperativity in DNA.

In another set of experiments\cite{directionalasymmetryjacs,jacssequencedependence}, the average unzipping times for a single molecule double-stranded DNA were found to be different depending upon the strand orientation during entry of the strand into the nanopore. This result is explained in those papers using the analogy of  a ``christmas tree'' moving through a hole, with the asymmetry of kinetics arising from the asymmetry of the tree structure. These experiments demonstrate the directionally asymmetric response of base-pair lifetime to nanopore probe. It is thus not unreasonable to assume that the bonding state of the left and right neighboring hydrogen bonds could similarly influence the lifetime of the middle hydrogen bond asymmetrically. In another experiment, even though the thermodynamic stabilities of the two sequences $5' \hyp(AT)_6(GC)_6\hyp 3'$ and $5' \hyp(GC)_6(AT)_6\hyp 3'$ are nearly the same, their unzipping kinetics have been shown to differ by orders of magnitude\cite{sequenceorderunzip}, suggesting that thermodynamics alone cannot explain the sequence functionality, and directionally asymmetric kinetic influences must be included. Unzipping kinetics of a DNA hairpin has been shown\cite{orientationdiscrimination} to strongly depend upon orientation of the terminal base-pairs. In another experiment\cite{duplexzippering}, adding the same four-nucleotide sequence to the $5'$ end of a longer sequence and to the $3'$ end of the same sequence resulted in significantly different zippering kinetics, with the effects on kinetics due to secondary structure formation explicitly ruled out.

The experiments cited above strongly suggest the presence of asymmetric cooperativity in DNA, which makes perfect sense, given the evolutionary advantage it provides to autocatalytic heteropolymers.

\subsubsection*{An experimental prediction}
Here, we make an experimentally verfiable claim, which cannot be explained by, to our knowledge, the only model explicitly built to explain the differences in the unzipping rates of DNA from either ends\cite{lambdadnaevidencefull}. Within the picture we developed here, the rates of unzipping of the sequence $5' \hyp (C)_n \hyp 3'/3' \hyp (G)_n \hyp 5'$, at constant force, should be different depending on the end where unzipping begins, and the rate of unzipping from the left end should be faster than from the right end, as suggested by both the experiment on incorporation kinetics\cite{incorporationkinetics} and genomic studies on $GC$ skew\cite{SkewReviewRocha,anotherskewreview}. This hypothesized outcome cannot be explained by the model constructed in \cite{lambdadnaevidence,lambdadnaevidencefull}, since that model requires sequence asymmetry to explain unzipping asymmetry, whereas, the above sequence is homogeneous. It is important that the proposed experiment is done at near-equilibrium conditions, as has been done in \cite{lambdadnaevidencefull}, to distinguish between thermodynamic and kinetic influences on the forces required for unzipping from both ends.

\section*{Conclusion}

We have found, in our model of self-replication of hypothetical autocatalytic heteropolymers, that unequal kinetic influence of inter-strand hydrogen bonds on their left and right neighbors improves the replicative potential substantially. This improvement is due to the simultaneous satisfaction of two competing requirements of both long lifetime of inter-strand hydrogen bonds to assist in covalent bonding, and low kinetic barrier for easy formation and dissociation of hydrogen bonds to speed up replication. This broken-symmetry mechanism is shown to lead to strand directionality.


\subsection*{Acknowledgements:} 
Support for this work was provided by the Moffitt Physical Science and Oncology Network (PS-ON) NIH grant, $U54CA193489$.
 
We thank Gerald Joyce, Antonio Lazcano, Addy Pross, John Cleveland, Joel Brown and Robert Gillies for useful comments. HS thanks Artem Kaznatcheev , IMO faculty and post-doctoral associates for helpful discussions.



\clearpage

\printbibliography


\section*{Statement of Competing Financial Interests}
The authors declare no competing financial interests.

\end{document}